\documentclass[aps,prd,showpacs,amsmath,10pt,twocolumn,superscriptaddress,floatfix,nofootinbib,notitlepage]{revtex4} 

\usepackage[utf8]{inputenc}
\usepackage[english]{babel}
\usepackage{bm,color,xcolor,amsfonts,amsmath,amssymb,epsfig,esint}

\newcommand{\be}{\begin{equation}}
\newcommand{\ee}{\end{equation}}
\newcommand{\bea}{\begin{eqnarray}}
\newcommand{\eea}{\end{eqnarray}}
\newcommand{\beas}{\begin{eqnarray*}}
\newcommand{\eeas}{\end{eqnarray*}}
\newcommand{\ds}{\displaystyle}

\newcommand{\Br}{\mbox{Br}}

\newcommand{\zb}{Z_b^{\pm}(10610)}
\newcommand{\zbp}{Z_b^{\pm}(10650)}
\newcommand{\zc}{Z_{c}^{\pm}(3900)}
\newcommand{\zcp}{Z_{c}^{\pm}(4020)}
\def\vec#1{\boldsymbol{#1}}

\newcommand{\Omnes}{Omn\`{e}s }

\newcommand{\fzj}{\affiliation{Institute for Advanced Simulation, Institut f\"ur Kernphysik and J\"ulich Center for Hadron Physics, Forschungszentrum J\"ulich, D-52425 J\"ulich, Germany}}

\newcommand{\jsi}{\affiliation{Jozef Stefan Institute, Jamova 39, 1000, Ljubljana, Slovenia}}

\newcommand{\lis}{\affiliation{CeFEMA, Center of Physics and Engineering of Advanced Materials, Instituto Superior T{\'e}cnico, Av. Rovisco Pais, 1 1049-001 Lisboa, Portugal}}

\newcommand{\rub}{\affiliation{Institut f\"ur Theoretische Physik II, Ruhr-Universit\"at Bochum, D-44780 Bochum, Germany}}

\graphicspath{{figs.dir/}}

\begin{document}

\title{On the emergence of heavy quark spin symmetry breaking in heavy quarkonium decays}

\author{V.~Baru}
\rub 

\author{E.~Epelbaum}
\rub

\author{A.A.~Filin}
\rub

\author{C.~Hanhart}
\fzj

\author{A.V.~Nefediev}
\jsi\lis

\begin{abstract}
Heavy-quark spin symmetry (HQSS) implies that in the direct decay of a heavy quarkonium with spin $S$, only lower lying heavy quarkonia with the same spin $S$ can be produced. However, this selection rule, expected to work very well in the $b$-quark sector, can be overcome if multiquark intermediate states are involved in the decay chain, allowing for transitions to the final-state heavy quarkonia with a different spin $S^{\prime}$. In particular, the measured decays $\Upsilon(10860)\to \pi Z_b^{(\prime)} \to \pi\pi\Upsilon(nS)$ ($n=1,2,3$) and $\Upsilon(10860)\to \pi Z_b^{(\prime)} \to \pi\pi h_b(mP)$ ($m=1,2$) appear to have nearly equal strengths { which is conventionally explained by a simultaneous presence of both $S_{b\bar{b}}=0$ and $S_{b\bar{b}}=1$ components in the wave functions of the $Z_b$'s in equal shares. 
Meanwhile, the destructive interference between the contributions of the $Z_b$ and $Z_b'$ to the decay amplitude for a $\pi\pi h_b$ final state kills the signal to zero in the strict HQSS limit.
In this paper, we discuss how the HQSS violation needs to be balanced by the narrowness of the $Z_b^{(\prime)}$ states in the physical case, to allow
for equal transition strengths into final states with different total heavy quark spins,
 and how spin symmetry is restored as a result of a subtle interplay of the scales involved, when the mass of a heavy quark becomes infinite. 
Moreover, we demonstrate how similar branching fractions of the decays
into $\pi\pi h_b$ and $\pi\pi\Upsilon$ can be obtained and how the mentioned HQSS breaking can be reconciled with the dispersive approach to the $\pi\pi/ K\bar K$ interaction in the final state and matched with the low-energy chiral dynamics in both final states.
}
\end{abstract}

\pacs{14.40.Rt, 11.55.Bq, 12.38.Lg, 14.40.Pq}


\maketitle


\section{Introduction}

The spectroscopy of hadronic states containing heavy quarks remains one of the fastest developing and most intriguing branches of studies of the strong interaction. 
Many new states have been discovered in the spectrum of charmonium and bottomonium, which do not fit into the quark model scheme and qualify as exotic states. For example, the states 
$\zb$, $\zbp$ \cite{Belle:2011aa}, $\zc$ \cite{Ablikim:2013mio,Liu:2013dau}, $\zcp$ \cite{Ablikim:2013wzq}, $Z^\pm(4430)$~\cite{Choi:2007wga,Mizuk:2009da,Chilikin:2013tch,Aaij:2014jqa} 
are charged 
and decay into final states containing a heavy quark $Q$ and its antiquark $\bar{Q}$ plus light hadrons. Therefore, their minimal quark content is four quarks. The interested reader can find a comprehensive overview of the current experimental and theoretical status of the exotic hadrons with heavy quarks in the dedicated review papers, for example, in Refs.~\cite{Lebed:2016hpi,Esposito:2016noz,Ali:2017jda,Guo:2017jvc,Olsen:2017bmm,Liu:2019zoy,Brambilla:2019esw}. 

The bottomoniumlike states $\zb$ and $\zbp$ (for brevity in what follows often referred to as $Z_b$ and $Z_b'$, respectively) are an ideal
laboratory to get a better understanding of exotic states, since they exist as two resonances with the same $J^{PC}=1^{+-}$~\cite{Collaboration:2011gja},
split by only about 45 MeV and are seen simultaneously in several modes. Specifically, the Belle Collaboration observed them as distinct peaks 
\begin{itemize}
\item[(i)] in the invariant mass distributions of the $\pi^\pm\Upsilon(nS)$ ($n=1,2,3$) and $\pi^\pm h_b(mP)$ ($m=1,2$) subsystems in the dipion transitions from the vector bottomonium $\Upsilon(10860)$ \cite{Belle:2011aa} and 
\item[(ii)] in the elastic $B\bar{B}^{*}$\footnote{Hereinafter a properly normalised $C$-odd combination of the $B\bar{B}^*$ and $\bar{B}B^*$ components is understood.} 
and $B^{*}\bar{B}^{*}$ channels in the decays $\Upsilon(10860)\to\pi B^{(*)}\bar{B}^*$ with the dominant branching fractions \cite{Adachi:2012cx,Garmash:2015rfd}. 
\end{itemize}

 In addition, the $Z_b(10610)$ and $Z_b(10650)$ states could in principle be experimentally searched for (and hopefully seen) in the $\rho \eta_b$ channel. More specifically, because of a large $\rho$-meson width, the relevant channel to look at would be $\Upsilon(10860)\to \pi Z_b^{(\prime)}\to\pi (\pi\pi)_{I=1}\, \eta_b$, which is however difficult experimentally because of a potentially large background related to the presence of the neutral pion in the final state. Meanwhile, no data for the given channel exist to date. On the other hand, the measured branching fractions $\Upsilon(10860)\to(\pi^+\pi^-\pi^0)_{\rm non-\omega}\, \chi_{bJ}$ ($J=1,2$) are by an order of magnitude suppressed 
\cite{Zyla:2020zbs} relative to the $\pi\pi\Upsilon(nS)$   and $\pi\pi h_b(mP)$  hidden-flavour decay channels considered in this work. One may therefore expect that the analogous branching for the $\pi (\pi \pi)_{I=1}\, \eta_b$ final state is also negligibly small. The fact that the $\rho$ and $\chi_{bJ}$ are in a relative $P$ wave in contrast to the $S$-wave final state $\rho\eta_b$ should not affect this estimate since both channels  open far away from the relevant energy range near the $B^{(*)}\bar{B}^*$ thresholds, so the centrifugal barrier suppression should not be operative any longer. Also, the measured elastic and inelastic branching fractions for $\Upsilon(10860)$ listed in the items (i) and (ii) above leave only very little room to other possible contributions.

The two most prominent explanations for the $Z_b$'s claimed to be consistent with the data are provided by a tetraquark model  
and a hadronic molecule picture,  see, e.g.,  Refs.~\cite{Esposito:2016noz,Ali:2017jda,Guo:2017jvc,Brambilla:2019esw}  for review articles and references therein. 
A review of the sum rules approach to the exotic states with heavy quarks and relevant references on the subject can be found in Ref.~\cite{Albuquerque:2018jkn}. It should be noted that a particularly close location of the $Z_b$'s to the thresholds of the $B\bar{B}^*$ and $B^*\bar{B}^*$ channels  
provides a strong hint in favour of their molecular interpretation. 

Both the $\zb$ and $\zbp$ contain a heavy $b\bar{b}$ pair, so it is commonly accepted that 
heavy-quark spin symmetry (HQSS) should 
be realised to high accuracy in these systems and indeed, as will be reviewed in Sec.~\ref{general}, 
 HQSS is able to explain naturally the interference pattern in the channels 
 $Z_b^{(\prime)}\to\pi\Upsilon(nS)$ and $Z_b^{(\prime)}\to\pi h_b(mP)$ \cite{Bondar:2011ev}.
 
{In Sec.~\ref{transitions}
we demonstrate how the experimental observation
that $\Br[\Upsilon(10860)\to \pi\pi h_b(mP)]$ is qualitatively similar or actually even larger than $\Br[\Upsilon(10860)\to\pi\pi\Upsilon(nS)]$
\cite{Zyla:2020zbs}
can be reconciled with HQSS and its violation. 
It appears as a result of a subtle interplay of the scales involved in
the system --- most relevant being the relationship
between the widths of the $Z_b$ states and their mass difference. Also, we discuss the restoration of spin symmetry
as the heavy quark mass is
taken to be infinitely large.}
 
Section~\ref{dispint} is devoted to the inclusion of the $\pi\pi/ K\bar K$ interaction in the final state. In particular, in
Refs.~\cite{Chen:2015jgl,Chen:2016mjn,Baru:2020ywb} it is explained how the $\pi\pi/ K\bar K$ final state interaction (FSI) can be included in the spin conserving heavy meson decays $\Upsilon\to\pi\pi\Upsilon'$ by means of twice subtracted dispersion integrals. 
Here the structure of the subtraction terms can be fixed by the chiral structure of the transition amplitudes,
giving rise to $\Upsilon\to \pi\pi\Upsilon'$ contact interactions.
While the chiral structure for
 the transitions of the kind $\Upsilon\to \pi\pi h_b$ is the same, such
counter terms violate HQSS and thus are expected to be strongly suppressed quantitatively. How 
this pattern can be reconciled with the $S$-wave $\pi\pi/ K\bar K$ FSI also for this kind of transitions is discussed in Sec.~\ref{dispint}. We summarise in Sec.~\ref{summary}.

\section{Spin wave functions}
\label{general}

The key that allows for a change of the heavy-quark spin in the presence of multiquark states
lies in the fact that the heavy-quark spins get rearranged in the assumed compact building blocks.
Indeed, let us stick to the strict HQSS limit of an infinite $b$-quark mass, $m_b\to\infty$ (quantities in this limit will be labelled by the superscript $^{(0)}$), and, following Ref.~\cite{Bondar:2011ev}, assume for illustration a molecular substructure for the $Z_b^{(0)}$ and $Z_b^{(0)\prime}$ states 
\be
Z_b^{(0)i} \sim B\bar{B}^*_i -\bar{B}B^*_i,\quad Z_b^{(0)\prime i} \sim i\epsilon_{ijk} B^*_j\bar{B}^*_k.
\label{Zbs}
\ee
Then, taking the spin wave functions of the $B$ and $B^*$ mesons in the form 
\begin{equation}
B=\psi_{\bar q}^\dagger \chi_b ,\quad B^*_i= \psi_{\bar q}^\dagger\sigma^i \chi_b,
\end{equation}
one arrives at the following spin structure of the $I^G(J^P)=1^+(1^+)$ molecular states
\begin{eqnarray}
Z_b^{(0)i} &\sim& ( \psi_{\bar q_1}^\dagger \chi_b)(\chi_{\bar b}^\dagger \sigma^i \psi_{q_2}){+}
( \psi_{\bar q_1}^\dagger \sigma^i \chi_b)(\chi_{\bar b}^\dagger \psi_{q_2}),\nonumber \\[-2mm]
\label{Zbs2}\\[-2mm]
Z_b^{(0)\prime i} &\sim& i\epsilon^{ijk} ( \psi_{\bar q_1}^\dagger \sigma^j \chi_b)(\chi_{\bar b}^\dagger \sigma^k \psi_{q_2}).\nonumber
\end{eqnarray}
Finally, via a Fierz rearrangement, these structures can be rewritten as
\begin{eqnarray}
Z_b^{(0)i} &\sim& ( \psi_{\bar q_1}^\dagger \psi_{q_2})(\chi_{\bar b}^\dagger \sigma^i \chi_b){+}
( \psi_{\bar q_1}^\dagger \sigma^i \psi_{q_2})(\chi_{\bar b}^\dagger \chi_b),\nonumber \\[-2mm]
\label{Zbs3}\\[-2mm]
Z_b^{(0)\prime i} &\sim& ( \psi_{\bar q_1}^\dagger \psi_{q_2})(\chi_{\bar b}^\dagger \sigma^i \chi_b){-}
( \psi_{\bar q_1}^\dagger \sigma^i \psi_{q_2})(\chi_{\bar b}^\dagger \chi_b),\nonumber
\end{eqnarray}
where the combinations $(\chi_{\bar b}^\dagger \chi_b)$ and
$(\chi_{\bar b}^\dagger \sigma^i \chi_b)$ correspond to the states
with $S_{b\bar{b}}=0$ and $S_{b\bar{b}}=1$, respectively. Therefore,
relations (\ref{Zbs3}) entail two important messages\footnote{Because
 of the coupled-channel transitions, the physical $Z_b$'s appear to
 be dynamical mixtures of the basis states introduced in
 Eq.~(\ref{Zbs}). Thus, these messages should be regarded as being originated from the strict HQSS.}
\begin{itemize}
\item in the strict HQSS limit both $Z_b$ states carry spin-0 and spin-1 components with equal weight and
\item {the spin-0 and spin-1 components}  appear with different relative signs in the two states.
\end{itemize}

{Importantly, as} demonstrated in Ref.~\cite{Bondar:2011ev}, the above {nontrivial} interference pattern is {indeed} consistent with the data, where the two $Z_b$ peaks appear to be equally strong in the decay chains 
\bea
\Upsilon(10860)&\to& \pi Z_b^{(\prime)}\to \pi\pi \Upsilon(nS),\quad n=1,2,3,\nonumber \\[-2mm]
\label{chains}\\[-2mm]
\Upsilon(10860)&\to& \pi Z_b^{(\prime)}\to \pi\pi h_b(mP),\quad m=1,2,\nonumber
\eea
with a constructive interference in the former and a destructive interference in the latter case.
{This underlines the fact that, although in general coupled-channel transitions mix the basis states (\ref{Zbs}), nevertheless, both wave functions of the physical $Z_b$'s still contain the $S_{b\bar{b}}=0$ and $S_{b\bar{b}}=1$ components in comparable shares.}
The same pattern emerges in the tetraquark picture proposed in Ref.~\cite{Maiani:2014aja}, where
the driving subclusters are compact diquarks and anti-diquarks. In contrast to the two scenarios just
sketched, it should be stressed that without {sizable heavy-quark spin-0 and spin-1 clusters} within the $Z_b$ states the observed pattern 
is unexplained and appears to be very unnatural. 

In the limit of the heavy quark mass going to infinity, the heavy-quark spin needs to be conserved. It is thus interesting to investigate, 
how this limit is reached in the scenario outlined above. It turns out that what controls the importance of the $h_b$ final states is the
ratio of the $Z_b$'s mass splitting, which violates HQSS and thus vanishes in the heavy quark limit, and the $Z_b$'s widths, which survive the
heavy quark limit. This will be discussed in detail below. 

\section{Spin-flip and spin-conserving transitions}
\label{transitions}

In this section we study the transition amplitudes for the reactions $\Upsilon\to \pi\pi\Upsilon'$ and $\Upsilon\to\pi\pi h_b$ via two $Z_b$ states and discuss (i) 
the scales that make it possible for
the two amplitudes
to have signals of a similar size for the physical masses of the heavy states involved and (ii) how the heavy-quark limit can be restored. 

Following Ref.~\cite{Bondar:2011ev}, the transition amplitudes read
\be
{\cal M}(\Upsilon\to\pi\pi\Upsilon')=C_{\Upsilon'} \left(\vec \Upsilon\cdot\vec \Upsilon^{\prime*}\right)\omega_+\omega_-[G_+(\omega_+)+G_+(\omega_-)]
\label{amplA}
\ee
and
\bea
{\cal M}(\Upsilon\to \pi\pi h_b)&=&C_{h_b}\left(\vec \Upsilon{\cdot} \left[\vec p_-\times\vec h_b^*\right]\right)\omega_+ G_-(\omega_+)\nonumber\\[-2mm]
\label{amplB}\\[-2mm]
&+&C_{h_b}\left(\vec \Upsilon{\cdot} \left[\vec p_+\times\vec{h}_b^*\right]\right)\omega_- G_-(\omega_-),\nonumber
\eea
where $\vec{\Upsilon}$, $\vec{\Upsilon}'$, and $\vec{h}_b$ are the
polarisation vectors of the $J^P=1^\pm$ states while
$C_{\Upsilon'}$, $C_{h_b}$ are normalisation constants. 
The Green's functions
$G_\pm$ describe the contribution of the $Z_b$'s in the intermediate state. The pion 4-momenta are taken in the form
\be
p_\pm=(\omega_\pm,\vec{p}_\pm).
\ee

Importantly, the relations between the spin wave functions (\ref{Zbs3}) imply that
\be
G_\pm=G_1\pm G_2,
\label{Gpm}
\ee
where $G_1$ and $G_2$ denote the individual propagators for the $Z_b$ and $Z_b'$ in the intermediate state, respectively. 
The nonrelativistic Green's functions for the $Z_b$'s read
\bea\label{Zs}
G_n(\omega_\xi)=\left(M_5-\omega_\xi-m_{Z_n}+\frac{i}{2}\Gamma_{Z_n}\right)^{-1},\nonumber
\eea
where $n=1,2$ and $\xi=\pm$. Further, $M_5$ stands for the mass of the
$\Upsilon(10860)$ and both $Z_b$'s are treated as unstable particles
with the masses $m_{Z_1}=m_z$ and $m_{Z_2}=m_z'$ and the widths
$\Gamma_{Z_1}=\Gamma_z$ and $\Gamma_{Z_2}=\Gamma_z'$, respectively. {These widths should be understood as visible widths of the $Z_b^{(\prime)}$ peaks in the experimental line shapes. }
Moreover, following Refs.~\cite{Bondar:2011ev,Voloshin:2011qa,Mehen:2011yh}, we assume $\Gamma_z'=\Gamma_z$. Here we emphasise that a quantitative treatment of the experimental line shapes
requires a coupled-channel effective-field-theory (EFT) based
framework manifestly consistent with unitarity, as given in
Refs.~\cite{Wang:2018jlv,Chen:2016mjn}, {in which the widths 
are generated dynamically}. However, for the purposes formulated above, that is, for studying a nontrivial interplay of various scales, the use of the simple analytic model of Ref.~\cite{Bondar:2011ev} is sufficient. 

Thus it is easy to find for the Green's functions $G_\pm(\omega_\xi)$ that
\bea
G_+(\omega_\xi)&=&\frac{1}{M_5-\omega_\xi-m_z+\frac{i}{2}\Gamma_z}+\frac{1}{M_5-\omega_\xi-m_z'+\frac{i}{2}\Gamma_z}\nonumber\\
 &=&\left(\frac{4}{\Gamma_z}\right)\frac{x_\xi+i}{(x_\xi+i)^2-r^2}\equiv\left(\frac{4}{\Gamma_z}\right)f_+(x_\xi,r),\label{fp}\\
G_-(\omega_\xi)&=&\frac{1}{M_5-\omega_\xi-m_z+\frac{i}{2}\Gamma_z}-\frac{1}{M_5-\omega_\xi-m_z'+\frac{i}{2}\Gamma_z}\nonumber\\
 &=&\left(\frac{4}{\Gamma_z}\right)\frac{-r}{(x_\xi+i)^2-r^2}\equiv \left(\frac{4}{\Gamma_z}\right)f_-(x_\xi,r),\label{fm}
\eea
where we introduced two dimensionless parameters,
\be
r=\frac{m_z'-m_z}{\Gamma_z}
\label{rdef}
\ee
and
\be
x_\xi=\frac{2}{\Gamma_z}(M_5-\bar{m}_z-\omega_\xi),
\label{xdef}
\ee
with the average mass $\bar{m}_z=(m_z+m_z')/2$.

To understand the behaviour of the amplitude in the heavy-quark limit, we need to discuss the scaling of
its different ingredients as functions of the heavy-quark mass $m_b$. 
First of all, it is easy to see that the width $\Gamma_z$ is (approximately)
independent of the heavy-quark mass. Indeed, the contributions from the inelastic (hidden-bottom) and elastic (open-bottom) channels to $\Gamma_z$ behave as
\bea
\Gamma_z^{\mbox{\scriptsize (in)}}&=&\frac{g_{\rm in}^2}{8\pi m_z^2}p_\pi={\cal O}(m_b^0),\nonumber\\[-2mm]
\\[-2mm]
\Gamma_z^{\mbox{\scriptsize (e)}}&=&\frac{g_{\rm e}^2}{8\pi m_z^2}p_B={\cal O}(m_b^0),\nonumber
\eea
where it was used that $g_{\rm in}\propto (\sqrt{m_b})^2\propto m_b$ due to the relativistic normalisation of the heavy fields, $g_{\rm e}^2\propto m_b^{3/2}$ as the coupling of a {compound state} to its constituents {\cite{Landau}}, and the pion and $B$-meson momenta scale with the heavy mass as $p_\pi={\cal O}(m_b^0)$ and $p_B\propto\sqrt{m_b}$, respectively. 

Then, according to the heavy-quark effective theory, the mass of an $S$-wave heavy-light $B^{(*)}$ meson scales as
\be
m_{B^{(*)}}=m_b+\bar{\Lambda}-\frac{\lambda_1+d_H\lambda_2}{m_b}+\cdots,
\label{mB}
\ee
where $\bar{\Lambda}\simeq\Lambda_{\rm QCD}$ is a universal parameter related to the light-quark dynamics while the $m_b$-independent constants $\lambda_1$ and $\lambda_2$ parametrise the contribution of the spin-dependent interactions with
\be
d_H=\left\{
\begin{array}{ll}
+3,&{\rm for}\;0^-\;{\rm state},\\
-1,&{\rm for}\;1^-\;{\rm state}.
\end{array}
\right.
\ee
The ellipsis in Eq.~(\ref{mB}) denotes terms of a higher order in the $1/m_b$ expansion. Using that the binding energies of $Z_b$ and $Z_b'$ are the same in the HQSS limit \cite{Bondar:2011ev,Voloshin:2011qa,Mehen:2011yh}, it is easy to see, therefore, that
\be
m_{z'}-m_z\simeq m_{B^*}-m_B={\cal O}\left(\frac{1}{m_b}\right),\quad
\bar{m}_z\approx 2m_b.
\label{mzz}
\ee

Finally, in the mass difference $M_5-\bar{m}_z$ the leading term $2m_b$ is cancelled, so that this difference tends to a constant in the limit $m_b\to\infty$.
Accordingly the parameter $x_\xi$ gets $m_b$ independent. Meanwhile, the parameter $r$ introduced in Eq.~(\ref{rdef}) scales as $r\propto 1/m_b$ and, therefore, can be used as a measure of the HQSS breaking effects. At the same time $r$ is the parameter that controls the strength of the dipion transitions from $\Upsilon(10860)$. Thus changing the pion energy $\omega_\xi$ (and accordingly changing $x_\xi$) allows one to scan through the $Z_b$ resonance structures to check their evolution with the variations of the parameter $r$. 

The physical masses and widths are taken from the Review of Particle Physics (RPP) by the PDG~\cite{Zyla:2020zbs} and read (in MeV) 
\be
m_z'-m_z=45,\quad \bar{m}_z=10630,\quad \Gamma_z=15,
\label{values}
\ee
where the width comes as an average between $\Gamma_{Z_b}$ and $\Gamma_{Z_b'}$.
Such parameters correspond to the physical value of the ratio
\be
r_{\rm phys}=3,
\label{rphys}
\ee
which is used as a reference point in what follows.

\begin{figure*}[t!]
\centering
\includegraphics[width=0.47\textwidth]{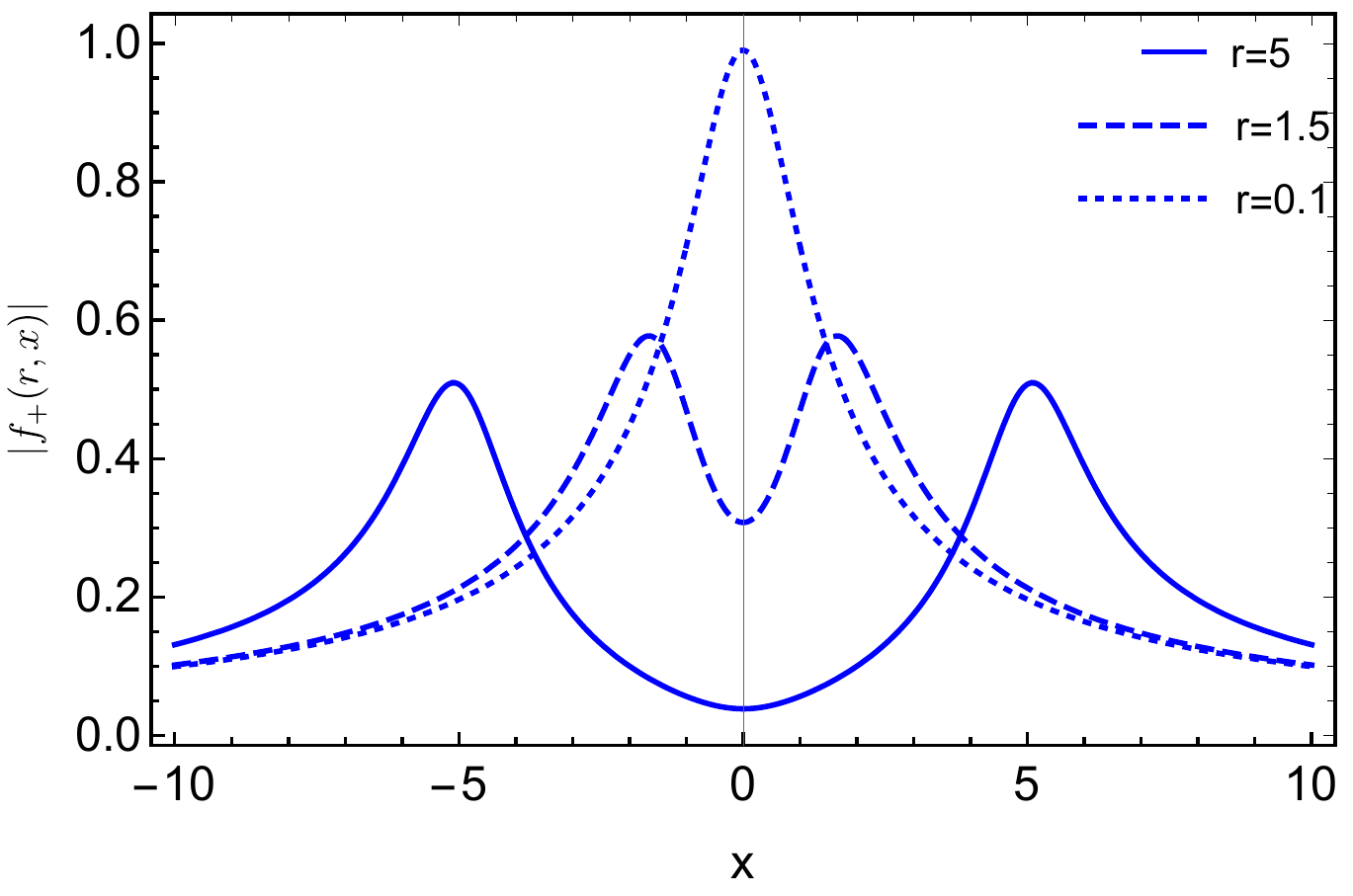}\hspace*{0.05\textwidth}\includegraphics[width=0.47\textwidth]{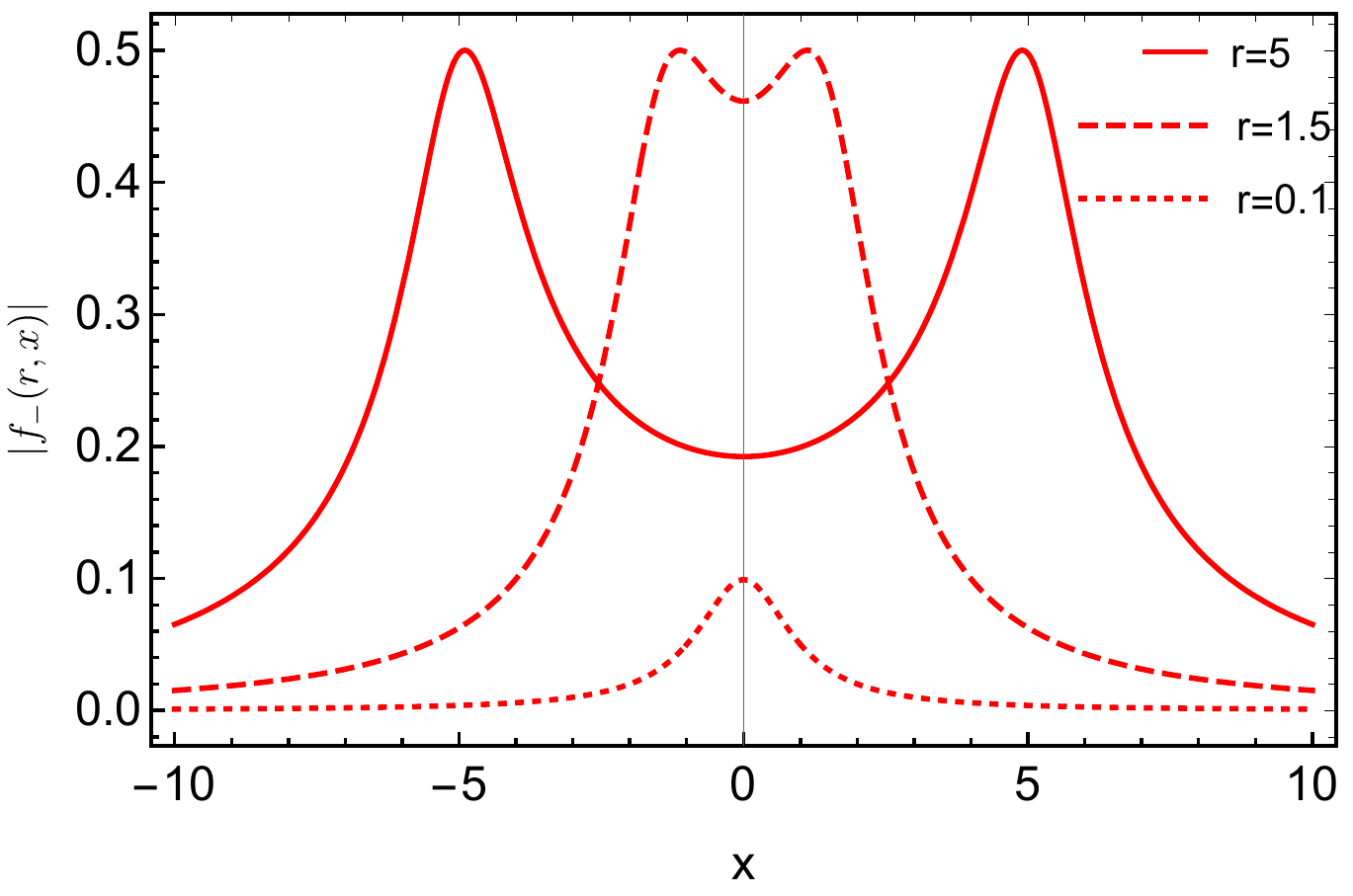}
\caption{The line shapes $|f_+(r,x)|$ (left panel) and $|f_-(r,x)|$ (right panel) as functions of $x$ {from Eq.~\eqref{xdef}. The function $f_+(r,x)$ ($f_-(r,x)$) shows a constructive (destructive) interference of the two $Z_b$ states as follows from Eq.~(\ref{fp}) (Eq.(\ref{fm})) for the $\pi\pi\Upsilon(nS)$ ($\pi\pi h_b(mP)$) final state.}
The results { shown by the solid, dashed, and dotted lines are for the HQSS violating parameter $r=5$, $r=1.5$ and $r=0.1$, respectively. The strict HQSS limit corresponds to $r=0$.}
}
\label{fig:lineshapes1}
\end{figure*}

\begin{figure*}[t!]
\centering
\includegraphics[width=0.5\textwidth]{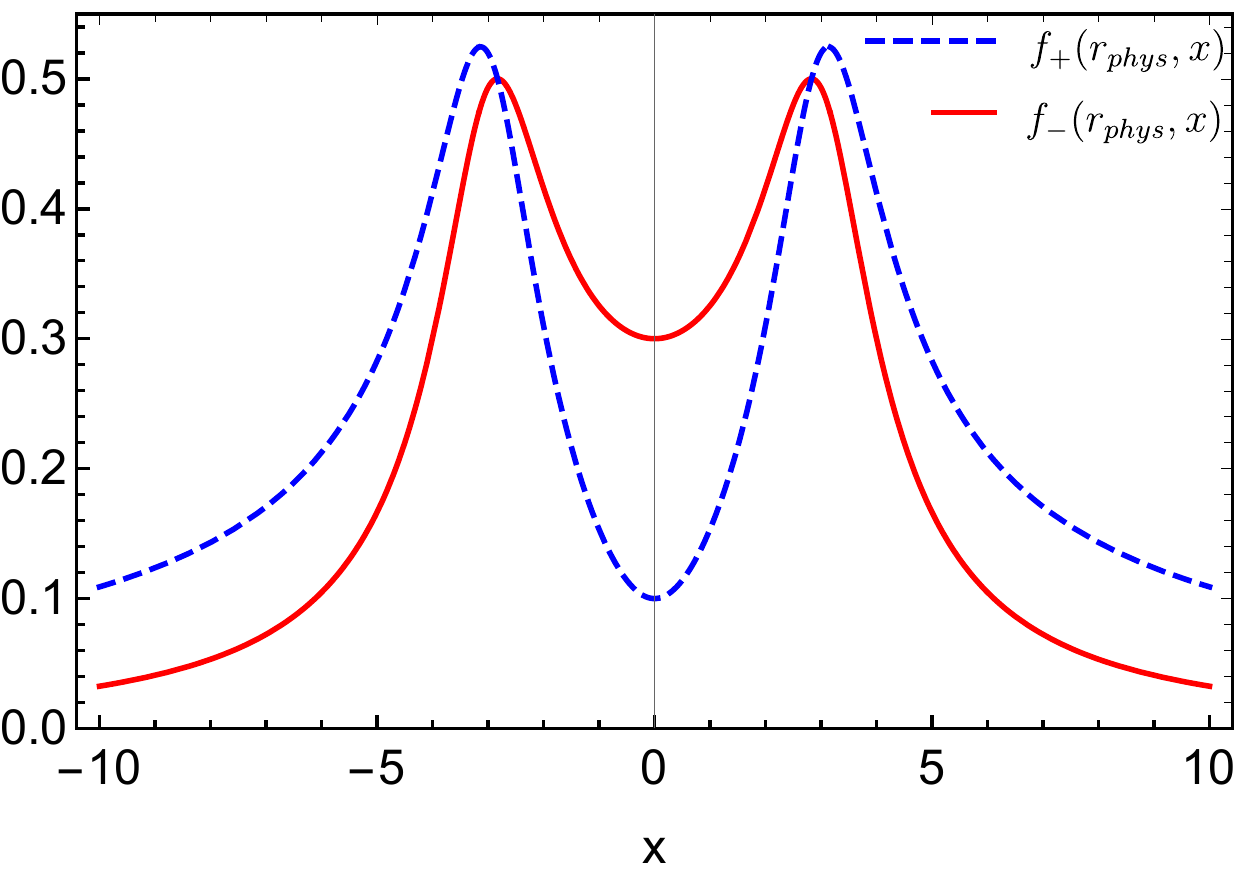}
\caption{The line shapes $|f_+(x,r)|$ (blue dashed curve) and $|f_-(x,r)|$ (red solid curve) as functions of $x$ which result from Eqs.~(\ref{fp}) and (\ref{fm}) for $r=r_{\rm phys}$ from Eq.~(\ref{rphys}). {See the caption of Fig.~\ref{fig:lineshapes1} for further details.}}
\label{fig:lineshapes2}
\end{figure*}

As an illustration we show in Fig.~\ref{fig:lineshapes1}
$|f_\pm(x,r)|$ as functions of $x$ ($x=x_+$ and $x=x_-$ result in
identical curves) for different values of $r$. It can be clearly seen
from this figure that the signals in the channels
$\Upsilon\to\pi\pi\Upsilon'$ and $\Upsilon\to \pi\pi h_b$, which are controlled
by the functions $f_\pm(x,r)$, respectively, demonstrate quite different patterns in the limit $r\to 0$. Indeed, while the constructive interference in the $\Upsilon'$ channel (function $f_+(x,r)$) provides the
sum of two peaks at $x=0$ for small values of $r$, the destructive interference in the $h_b$ channel (function $f_-(x,r)$) results in a severe damping of
the signal in this limit, namely,
\bea
|f_+(x=0,r)|&\ds\mathop{\to}_{r\to 0}& 1,\nonumber\\[-2mm]
\label{f_r_0}\\[-2mm]
|f_-(x=0,r)|&\ds\mathop{\approx}_{r\to 0}& r \to 0.\nonumber
\eea
Thus the HQSS indeed gets restored in the heavy quark limit, and the rate of this restoration is controlled
by the parameter $r$.

In the meantime, in the opposite regime of $r\gg 1$, in which the effect of the HQSS violation is balanced by the narrow width [see Eq.~\eqref{rdef}], both the peak positions and their strengths coincide for $f_+(x,r)$ and $f_-(x,r)$: 
\bea
&\ds x_{\rm peak}=\pm r + {\cal O}(r^{-1}),&\nonumber\\[-2mm]
\label{f_r_inf}\\[-2mm]
&\ds |f_\pm(x_{\rm peak},{r\to\infty})|=\frac12+{\cal O}(r^{-1}).&\nonumber
\eea 

It is therefore not surprising that already for $r=r_{\rm phys}$, the curves $|f_+(x,r)|$ and $|f_-(x,r)|$ demonstrate a noticeable similarity both in the shape and strength --- see Fig.~\ref{fig:lineshapes2}.

Finally, we study the behaviour of the differential rates for the dipion transitions from the $\Upsilon(10860)$. The doubly differential decay width for a three-body final state takes the form
\be
d\Gamma=\frac{1}{(2\pi)^3}\frac{1}{32M_5^3}\overline{|{\cal M}|^2}dm_{\pi\pi}^2dm_{\pi h_b/\pi\Upsilon'}^2,
\ee
where, depending on the process considered, $\overline{|{\cal M}|^2}$
is the decay amplitude (\ref{amplA}) or (\ref{amplB}) squared and
averaged over the polarisations of the vector (axial vector)
particles. Then, expressing
$m_{\pi h_b/\pi\Upsilon'}^2$
in terms of 
$x=x_+$ and performing the integration over the dipion invariant mass, it is straightforward to arrive at the differential rate in the form
\be
\frac{d\Gamma}{dx}=\frac{\Gamma_z}{(2\pi)^3 32M_5^2}\int_{(m_{\pi\pi}^2)_{\rm min}}^{(m_{\pi\pi}^2)_{\rm max}}\overline{|{\cal M}|^2}\;dm_{\pi\pi}^2.
\label{rates}
\ee

\begin{figure*}[t!]
\centering
\includegraphics[width=0.47\textwidth]{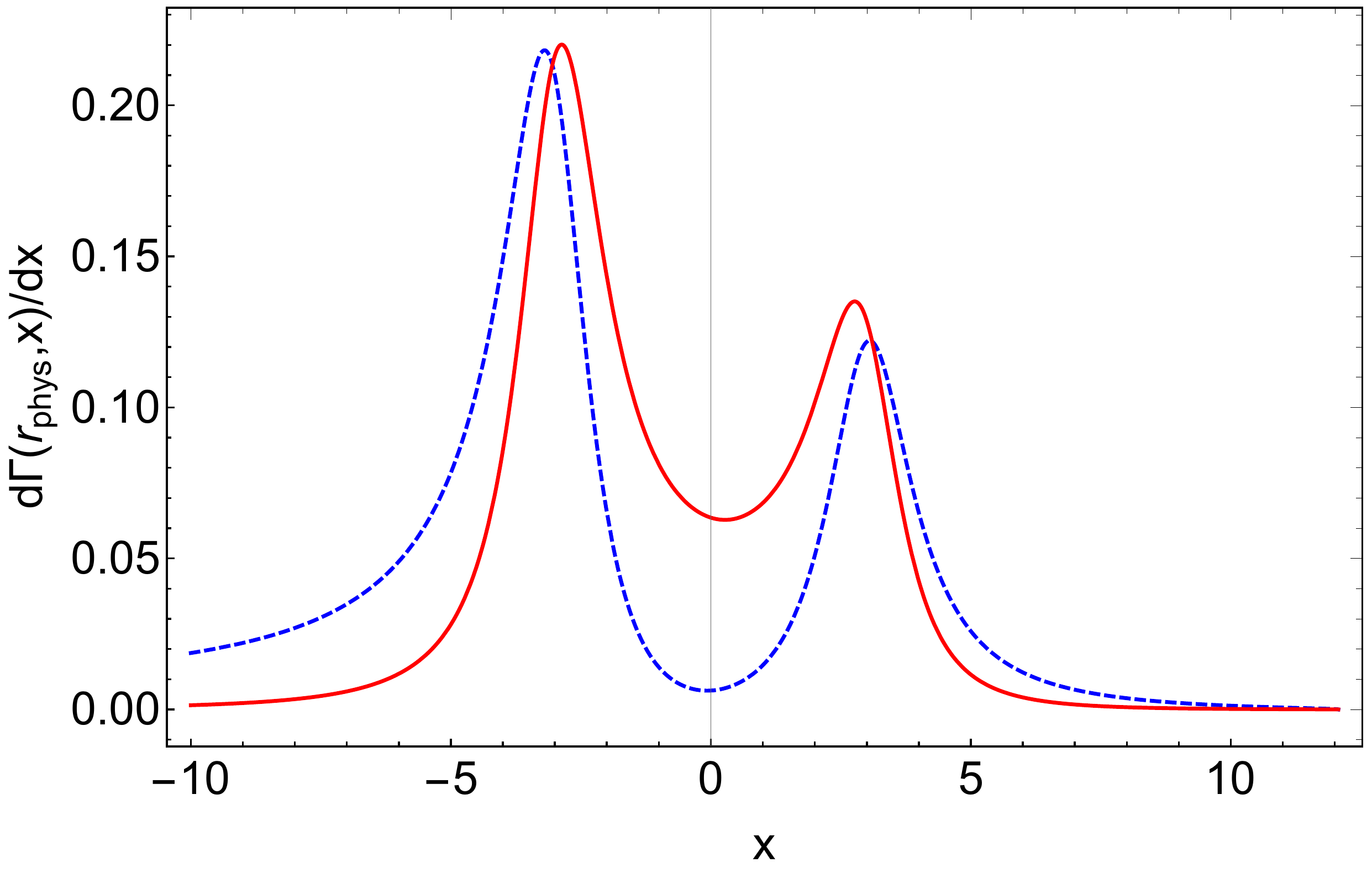}\hspace*{0.04\textwidth}\includegraphics[width=0.45\textwidth]{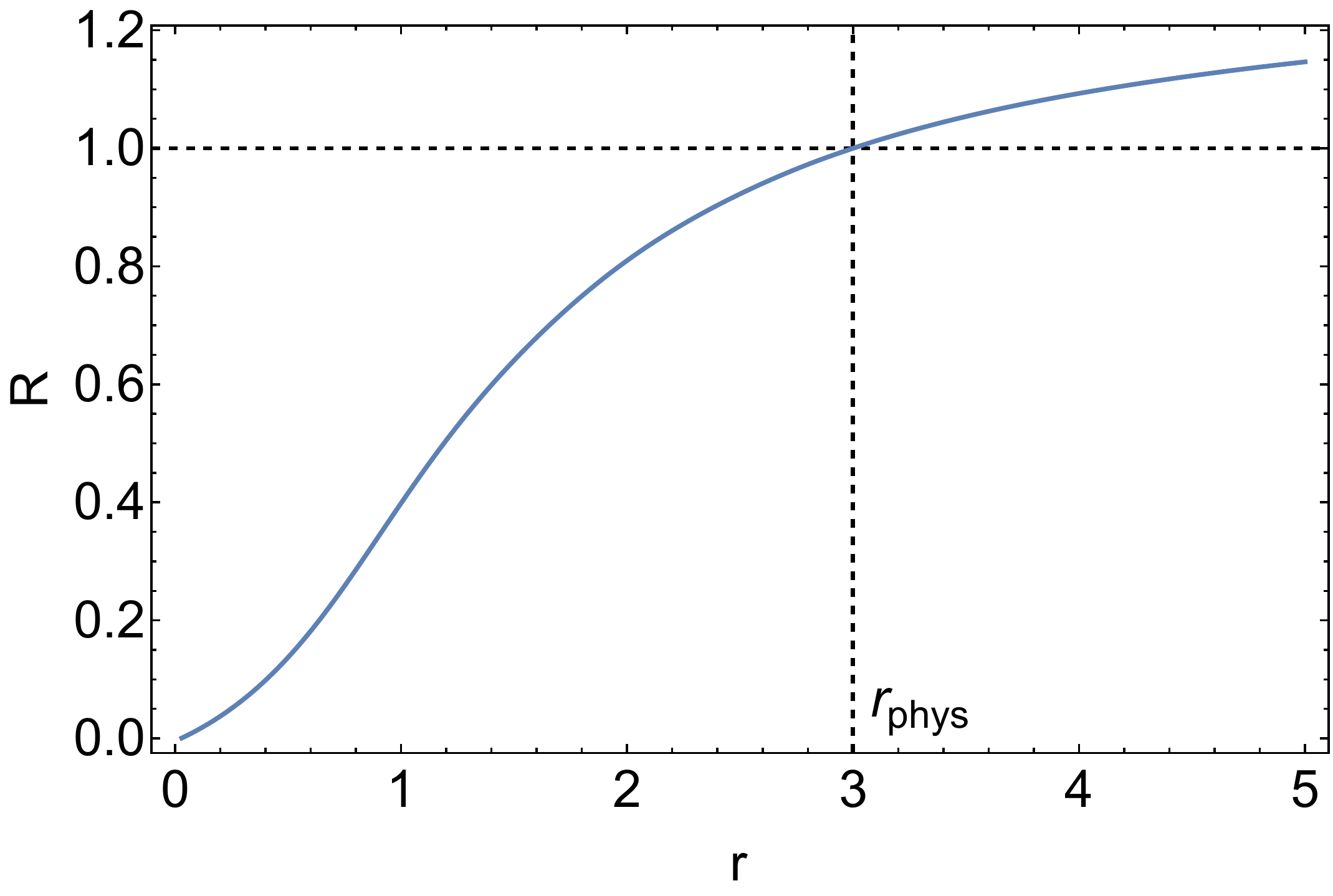}
\caption{Left panel: The differential rates (in arbitrary units) of the dipion transitions $\Upsilon(10860)\to\Upsilon(1S)\pi\pi$ (blue dashed line) and $\Upsilon(10860)\to \pi\pi h_b$ (red solid line), as defined in Eq.~(\ref{rates}) and normalised as given in Eq.~(\ref{eqbr}). Right panel: the ratio (\ref{ratio}) as a function of the parameter $r$. The dashed lines pinpoint the values $r_{\rm phys}\approx 3$ and $R(r_{\rm phys})=1$. To ease the comparison, in this figure,
the masses of both quarkonia in the final states, $\Upsilon'$ and $h_b$, are taken equal to 9.899 GeV (the mass of $h_b(1P)$~\cite{Zyla:2020zbs}).
}
\label{fig:lineshapes3}
\end{figure*}

For the presentation purposes we fix the ratio of the unknown overall factors ${\cal C}_{\Upsilon'}$ and ${\cal C}_{h_b}$ introduced in the amplitudes (\ref{amplA}) and (\ref{amplB}) such that the total decay widths, evaluated as the integrals from the differential width (\ref{rates}) over the region in the $x$ covering the $Z_b$ peaks, obey the constraint 
\bea
&&\Gamma[\Upsilon\to\pi Z_b^{(\prime)}\to\pi\pi\Upsilon'](r_{\rm phys})\hspace*{0.3\columnwidth}\nonumber\\[-2mm]
\label{eqbr}\\[-2mm]
&&\hspace*{0.3\columnwidth}\simeq \Gamma[\Upsilon\to\pi Z_b^{(\prime)}\to \pi\pi h_b](r_{\rm phys}),\nonumber
\eea
in qualitative agreement with the data \cite{Zyla:2020zbs}. 
The resulting normalised line shapes (\ref{rates}) are depicted (in arbitrary units) in the left panel of Fig.~\ref{fig:lineshapes3} from which one can see a noticeable similarity of the curves in the shape.

We also define the ratio
\be
R=\frac{\Gamma[\Upsilon\to \pi\pi h_b](r)}
{\Gamma[\Upsilon\to\pi\pi\Upsilon'](r)}
\label{ratio}
\ee
and plot it as a function of the parameter $r$ in the right panel of Fig.~\ref{fig:lineshapes3}. From this plot one can see that, in agreement with the considerations of this section, the probability of the decay $\Gamma[\Upsilon\to \pi\pi h_b]$ vanishes in the strict HQSS limit (at $r=0$) but then grows with $r$ to reach $\Gamma[\Upsilon\to \pi\pi\Upsilon']$ at the physical point (\ref{rphys}), as prescribed by the normalisation condition (\ref{eqbr}). 

Notice also that, alternatively, the limit $r\to 0$ can be reached by
increasing $\Gamma_z$ --- see the definition (\ref{rdef}) --- so that
the small $Z_b$ widths play an important role in explaining
the observed large branching fractions for the HQSS violating decays $\Upsilon(10860)\to \pi\pi h_b$.

\section{Inclusion of the pion interaction in the final state}
\label{dispint}

\subsection{General remarks}

In the previous section it was demonstrated how the explicit HQSS
breaking due to the $Z_b$ mass splitting could be balanced by their
narrow widths to provide similar branching fractions for the reactions
with and without the heavy-quark spin flip. Here we discuss
consequences of yet another peculiar property of the HQSS violation in
the reaction $\Upsilon(10860)\to \pi\pi h_b(mP)$ as compared with the
$\pi\pi\Upsilon(nS)$ final states. {Namely, the tree level amplitude of the
reaction $\Upsilon(10860)\to \pi\pi\Upsilon(nS)$ admits two
momentum-dependent chiral
contact terms (with two derivatives of the pion fields), 
which parameterize a short range interaction in
the system, already in the strict heavy-quark
limit~\cite{Mannel:1995jt,Chen:2015jgl,Chen:2016mjn,Baru:2020ywb}. These contact terms can be matched with two subtraction
constants needed to include the $\pi\pi/K\bar K$ FSI in this
reaction.} 
In contrast, the leading contact interactions in the decays $\Upsilon(10860)\to \pi\pi h_b(mP)$, while
having the same chiral structure,   violate HQSS and therefore should be strongly suppressed relative to those for the transitions
that conserve the heavy quark spin. In what follows, we discuss how
this fact can be reconciled with the fact that in the individual reactions $$\Upsilon(10860)\to \pi Z_b \to \pi\pi h_b(mP)$$ and $$\Upsilon(10860)\to \pi Z_b^{\prime} \to \pi\pi h_b(mP),$$ the same number of subtractions is required to render the dispersive integrals finite as in the channel with the $\pi\pi\Upsilon(nS)$ final state. Specifically, we demonstrate that the destructive interference of the two amplitudes in the $\pi\pi h_b$ channel (as opposed to the constructive interference for the $\pi\pi\Upsilon(nS)$ final state) provides a better convergence of the resulting dispersive integral, so that a smaller number of subtractions is required.

To bring a quantitative twist to this discussion, we employ the coupled-channel EFT-based approach developed for the $Z_b$ states in Refs.~\cite{Wang:2018jlv,Baru:2019xnh} with the $\pi\pi/K\bar K$ FSI
for the transition $ \Upsilon(10860)\to \pi\pi h_b(mP)$ to demonstrate that its role in this channel is indeed suppressed, in contrast to the $\pi\pi\Upsilon(nS)$ final states \cite{Baru:2020ywb}.

\subsection{Kinematics and notations}

The kinematics of the decays 
\be
\Upsilon(p_i)\to \pi^+(p_+)\pi^-(p_-)\Upsilon'/h_b(p_f),
\ee
is conventionally introduced through the Mandelstam invariants,
\be
s=(p_i+p_f)^2,\quad t=(p_f+p_+)^2,\quad u=(p_f+p_-)^2,
\ee
with
\be
p_i^2=m_i^2,\quad p_f^2=m_f^2,\quad p_+^2=p_-^2=m_\pi^2,
\ee
where $m_\pi$, $m_i$, and $m_f$ are the masses of the pion, $\Upsilon\equiv\Upsilon(10860)$, and $\Upsilon'\equiv\Upsilon(nS)$/$h_b\equiv h_b(mP)$, respectively. 
As always, one has
\be
s+t+u=m_i^2+m_f^2+2m_\pi^2.
\ee 
The chosen kinematics
 corresponds to considering the amplitude ${\cal M}(s,t,u)$ in a crossed channel, $\Upsilon(p_i)+\Upsilon'/h_b(p_f)\to\pi^+(p_+)+\pi^-(p_-)$.
 It proves convenient to pass over to the center of mass frame of the pions in the final state, so that their 4-momenta now read
\be
p_\pm=(\omega_\pi,\pm\vec{p}_\pi).
\ee
In the given kinematics, 
\bea
t(s,z)-m_z^2&=-&\frac12 \biggl(Y(s,m_z)-k(s)z\biggr),
\nonumber \\[-2mm]
\label{tandu}\\[-2mm]
u(s,z)-m_z^2&=-&\frac12 \biggl(Y(s,m_z)+k(s)z\biggr),
\nonumber
\eea
where
\be
Y(s,m_z)=s-m_i^2-m_f^2-2m_{\pi}^2+2m_z^2
\ee
and
\be
k(s)=\frac{1}{s}\sqrt{\lambda(s,m_i^2,m_f^2)\, \lambda(s,m_{\pi}^2,m_{\pi}^2)},
\ee
with $\lambda(m_1^2,m_2^2,m_3^2)$ for the standard K\"allen triangle function,
\be
\lambda(x,y,z)=x^2+y^2+z^2-2xy-2xz-2yz.
\label{lambda}
\ee
Then, for $z \equiv \cos \theta$
(here $\theta$ is the angle between the 3-momenta ${\bm p}_\pi$ and ${\bm p}_f$) one finds
\be
z=\dfrac{t-u}{k(s)}.
\label{z}
\ee

\subsection{Production amplitudes through $Z_b$'s}

\begin{figure*}[t!]
\centering
\includegraphics[width=0.45\textwidth]{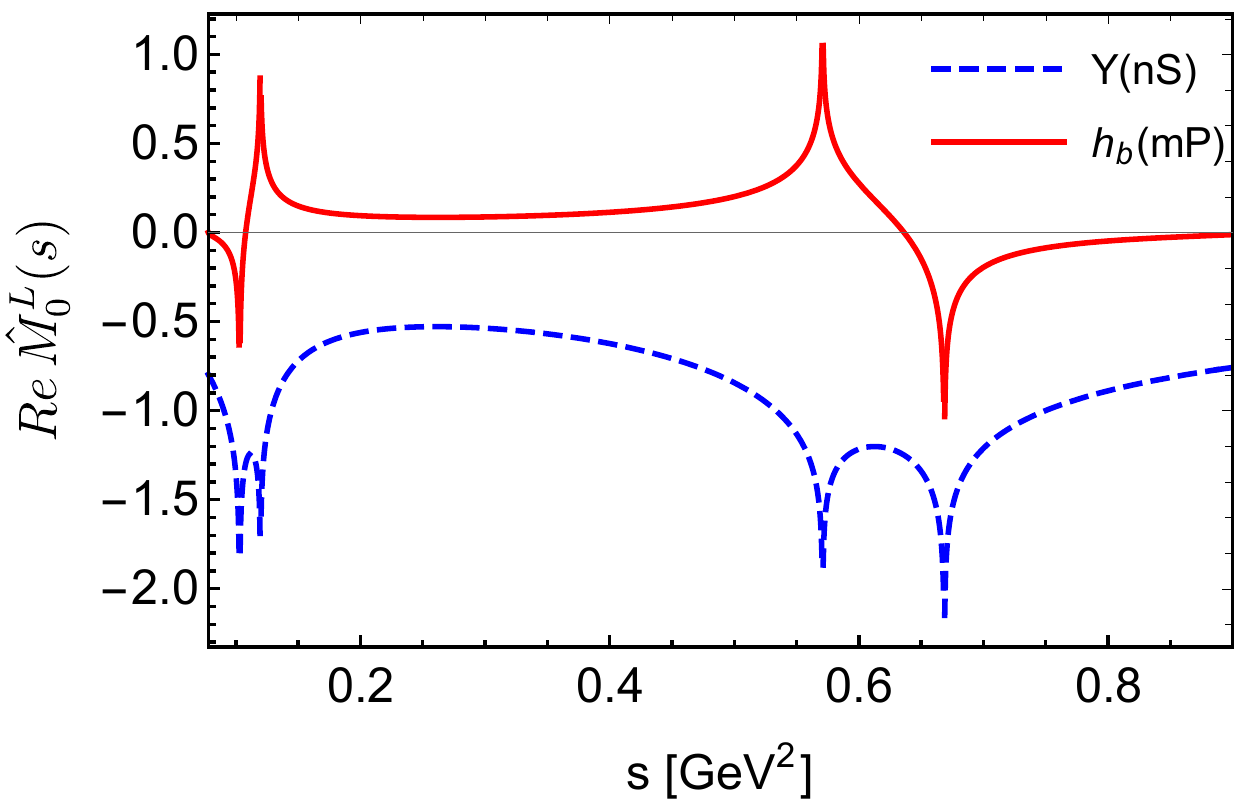}\hspace*{0.1cm}
\includegraphics[width=0.45\textwidth]{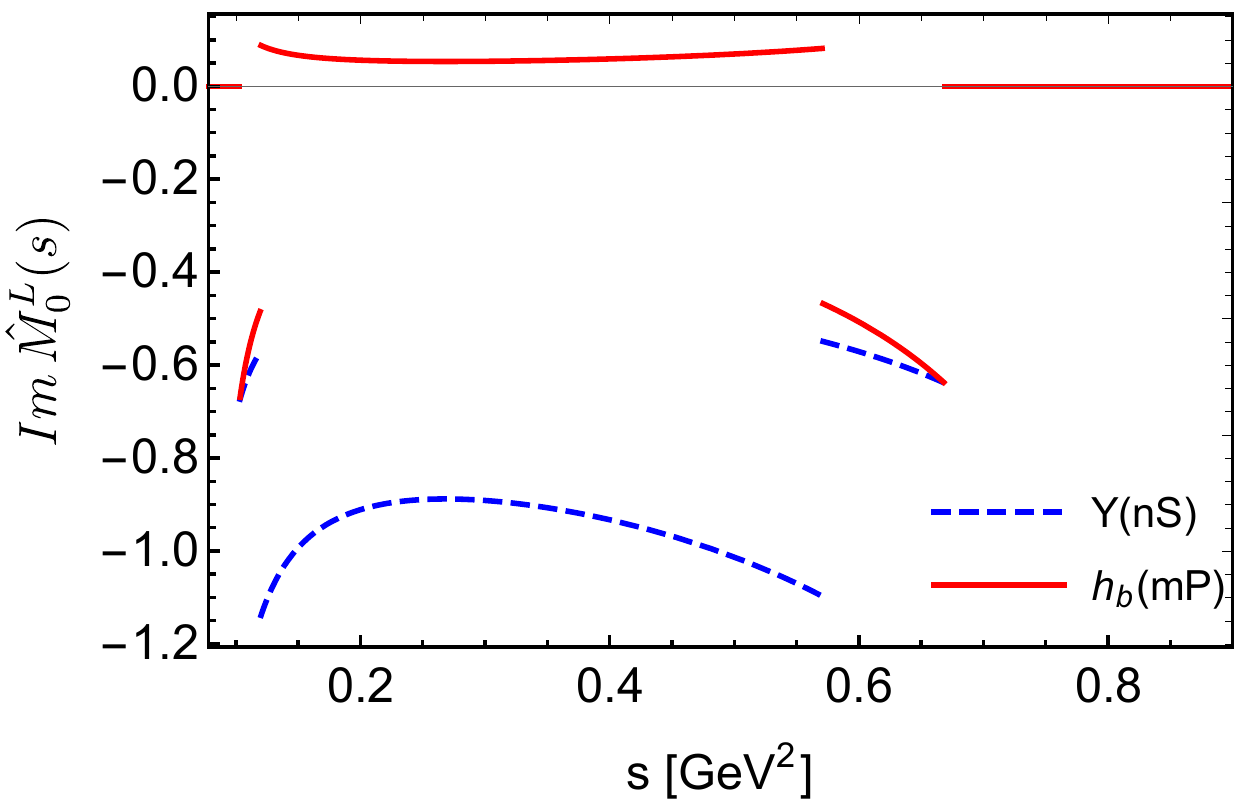}
\caption{{Comparison of} the $S$-wave-projected left-hand cut amplitudes via the $Z_b(10610)$ and $Z_b(10650)$ for the $\Upsilon$ (blue dashed lines) and $h_b$ (red solid lines) final states, defined by Eqs.~\eqref{MLUps} and \eqref{MLhb}, respectively. 
{The figure shows that 
because of 
a very different interference patterns between the two $Z_b$ contributions (see especially the left panel),  
the resulting signal for the $h_b$ final state is much smaller than the one in the $\Upsilon$ channel, and will be further suppressed after ``averaging'' in the dispersive integral \eqref{Idisp}.
}
To ease the comparison, in this figure, the masses of both quarkonia in the final states, $\Upsilon'$ and $h_b$, are taken equal to 9.899 GeV (the mass of $h_b(1P)$~\cite{Zyla:2020zbs}).}
\label{fig:ML}
\end{figure*} 

The amplitudes for the decays $\Upsilon(10860)\to \pi\pi\Upsilon(nS)$ and $\Upsilon(10860)\to \pi\pi h_b(mP)$, which proceed through the $Z_b$ and $Z_b'$, can now be written as 
\be
{\cal M}(\Upsilon\to\pi\pi\Upsilon')=C_{\Upsilon'}' \left(\vec \Upsilon\cdot\vec \Upsilon^{\prime*}\right) 
[G_+(t)+G_+(u)]
\label{MU}
\ee
and
\be
{\cal M}(\Upsilon\to \pi\pi h_b)=C_{h_b}'\left(\vec \Upsilon{\cdot} \left[\hat{\vec p}_\pi\times\vec h_b^*\right]\right)[G_-(t)-G_-(u)],
\label{Mh}
\ee
respectively, where $\hat{\vec p}_\pi = {\vec p}_\pi/ p_\pi$, $G_\pm$, as before, include the contributions from both $Z_b$ states
(see Eq.~(\ref{Gpm})), and the individual propagators of the $Z_b$'s read
\be
G_1(\tau)=\frac{1}{\tau-m_z^2},\quad G_2(\tau)=\frac{1}{\tau-m_{z'}^2},
\ee
with $\tau=t,u$. Here we absorbed all slowly varying functions of
momenta in the numerators into the normalization constants and
neglected the widths of the $Z_b$'s which do not play a role for the
argument. Indeed, as one can see from the right plot in
Fig.~\ref{fig:lineshapes3} (see also the discussion around
Eq.~\eqref{f_r_inf}), the function $R(r)$ is almost
constant at $r>r_{\rm phys}$ with the value of $r_{\rm phys}$ quoted in Eq.~(\ref{rphys}). In other words, to pinpoint the main difference between the asymptotic (large-$s$) behaviour of the functions $G_+$ and $G_-$, it is safe to take the limit of $r\to \infty$ achieved for $m_z'\neq m_z$ and $\Gamma_z\to 0$ --- see the definition of $r$ in Eq.~(\ref{rdef}). 
We also note that the effect of a finite width will be included in the full calculation performed in Sec.~\ref{sec:full} below.

A nontrivial difference in the production of the $\Upsilon'$ and $h_b$ states can be seen directly from the expressions for $G_\pm(\tau)$ which read
\bea
G_+(\tau)&=&
\frac{1}{\tau-m_z^2}+\frac{1}{\tau-m_z'^2}
=\frac{2\tau-m_z^2-m_z'^2}{(\tau-m_z^2)(\tau-m_z'^2)},\nonumber\\[0mm]
\label{amlsUh}\\[-2mm]
G_-(\tau)&=&
\frac{1}{\tau-m_z^2}-\frac{1}{\tau-m_z'^2}
=\frac{m_z^2-m_z'^2}{(\tau-m_z^2)(\tau-m_z'^2)}.\nonumber
\eea
First, one concludes that $G_-(\tau)$ has a milder UV behavior at large $\tau$ (and therefore also at large $s$, by virtue of Eqs.~\eqref{tandu}) than $G_+(\tau)$. Second, the result for $G_-(\tau)$
obviously vanishes in the strict HQSS limit, while $G_+(\tau)$ does not. 

The interaction of the pions in the final state of the reaction $\Upsilon(10860)\to \pi\pi\Upsilon(nS)$ was studied in detail in Ref.~\cite{Baru:2020ywb}, where
 a dispersive approach was developed for the Born amplitude 
\be
M_{\rm stable}^\Upsilon(t,u;m_z)=\frac{1}{t-m_z^2} +\frac{1}{u-m_z^2}.
\label{MBorn0}
\ee
In particular, the $S$-wave-projected amplitude with only left-hand cuts was defined as
\be
M_{\rm 0,stable}^{\Upsilon; L}(s,m_z)=\frac12\int_{-1}^1dz \, M_{\rm stable}^\Upsilon(t,u;m_z)
\label{MBorns}
\ee
and found in a closed form, including anomalous pieces. 

Given that the left-hand cut amplitude $M_0^L$ is known, the total amplitude including its part with the right-hand cuts due to the $\pi\pi$ and $K\bar K$ interaction in the final state can be restored dispersively as
\be
\hat{M}_0(s)=\hat{M}_0^L(s)+
\hat{\Omega}_0(s)\hat{I}_0(s),
\label{Mmultch}
\ee
with
\be
\hat{I}_0(s)=\frac{1}{\pi}\int_{4m_\pi^2}^\infty\, ds'\frac{\hat{\Omega}_0^{-1}(s')\hat{T}(s')\hat{\sigma}(s')\hat{M}_0^L(s')}{s'-s-i0},
\label{Idisp}
\ee
where $\hat{\Omega}_0(s)$ is the multichannel \Omnes matrix\footnote{Since only isoscalars are considered in this work, the isospin index of the \Omnes matrix is omitted.} with the hats indicating multicomponent objects (vectors and matrices), $\hat{\sigma}(s)=\text{diag}\{\sigma_\pi, \sigma_K\}$ is a diagonal matrix with 
$\sigma_P(s)=\sqrt{1-s_P^{\rm th}/s}$ and $s_P^{\rm th}$
denoting the threshold in the corresponding channel ($P=\pi,K$). In particular, $\hat{M}_0^L=\left([M_0^L]_{\pi\pi},[M_0^L]_{KK}\right)^T$. The $S$-wave meson-meson coupled-channel amplitude $\hat T$ can be parametrised by the $\pi\pi$ scattering phase shift $\delta(s)$ \cite{GarciaMartin:2011cn,Caprini:2011ky,Buettiker:2003pp,Dai:2014zta} as well as the absolute value and phase of the $\pi\pi \to K\bar K$ transition \cite{Buettiker:2003pp,Dai:2014zta}.
For the reaction $\Upsilon\to\pi\pi\Upsilon'$ the leading contribution comes from the $Z_b$ and $Z_b'$ states, and thus one finds
\be\label{MLUps}
\hat{M}_0^{\Upsilon; L}(s)=M_{\rm 0,stable}^{\Upsilon; L}(s,m_z)+ M_{\rm 0,stable}^{\Upsilon; L}(s,m_z').
\ee

Consider now the $S$-wave-projected amplitude for the $h_b$ channel,
\be
M_{\rm 0,stable}^{h_b; L}(s,m_z)=\frac12\int_{-1}^1zdz \, M_{\rm stable}^{h_b}(t,u;m_z),
\label{MBorns2}
\ee
where the factor $z$ under the integral comes from the $P$-wave between the pions and the $h_b$. According to Eq.~(\ref{Mh}), the Born amplitude reads
\be
M_{\rm stable}^{h_b}(t,u;m_z)=\frac{1}{t-m_z^2}-\frac{1}{u-m_z^2}.
\label{MBorn02}
\ee

Since $M_{\rm stable}^{h_b}(t,u;m_z)\propto (t-u)\propto z$, the projection integral (\ref{MBorns2}) is non zero and can be expressed in terms of $M_{\rm 0,stable}^{\Upsilon;L}$ from Eq.~(\ref{MBorns}) as
\bea
M_{\rm 0,stable}^{h_b;L}(s,m_z)= \frac12\int dz \left(\frac1{t-m_z^2}-\frac1{u-m_z^2}\right) z \nonumber\\
=\frac{1}{k(s)}\left(4+Y(s,m_z) M_{\rm 0,stable}^{\Upsilon;L}(s,m_z)\right).
\eea
Then, the full left-hand cut amplitude through the $Z_b$ and $Z_b'$ reads (\emph{cf.} Eq.~\eqref{MLUps})
\bea\nonumber
\hat{M}_0^{h_b; L}(s)&=&M_{\rm 0,stable}^{h_b; L}(s,m_z)- M_{\rm 0,stable}^{h_b; L}(s,m_z')\\ \nonumber
 &=&\frac{1}{\kappa(s)} 
 \left(Y(s,m_z) M_{\rm 0,stable}^{\Upsilon;L}(s,m_z)\right.\\
&&\hspace*{0.12\columnwidth} \left. -Y(s,m_z') M_{\rm 0,stable}^{\Upsilon;L}(s,m_z')\right). \label{MLhb}
\eea
In contrast to the $\pi \Upsilon/\pi h_b$ invariant mass distributions, where the absolute values of the amplitudes are relevant (see Sec.~\ref{transitions}),
the evaluation of the dispersive integral in Eq.~\eqref{Idisp} involves the Re and Im parts of the left-hand cut amplitude $\hat M^L_0(s)$ individually. 
In Fig.~\ref{fig:ML} we compare
the Re and Im parts of the $S$-wave-projected left-hand cut amplitudes defined in Eqs.~\eqref{MLUps} and \eqref{MLhb}
for the $\Upsilon$ and $h_b$ final states, respectively. Because of a different interference, the shapes of the amplitudes $\hat{M}_0^{\Upsilon; L}(s)$ 
and $\hat{M}_0^{h_b; L}(s)$ are very different, namely, while the contributions of the $Z_b(10610)$ and $Z_b(10650)$ add up constructively in $\hat{M}_0^{\Upsilon; L}(s)$,
they interfere destructively in $\hat{M}_0^{h_b; L}(s)$. It is therefore natural to expect that the net $\pi\pi/K\bar K$ FSI contribution to $\Upsilon(10860)\to \pi\pi h_b(mP)$, which violates HQSS, will be strongly suppressed.

It was argued in Ref.~\cite{Baru:2020ywb} that the asymptotic behavior of the $S$-wave projected left-hand cut amplitude $M_{\rm 0,stable}^{\Upsilon;L}$ is 
\be
\hat{M}_{\rm 0,stable}^{\Upsilon;L}\mathop{\propto}_{s\to\infty}\frac{\ln(s)}{s}.
\label{M0asympt}
\ee
Then,  one concludes from Eq.~\eqref{MLhb} that 
\be
\hat{M}_0^{h_b; L}(s) \mathop{\propto}_{s\to\infty} \left (m_z^2-m_z'^2\right)\frac{\ln(s)}{s^2},
\ee
that is, in line with the argument around Eq.~\eqref{amlsUh}, the transition amplitudes $\Upsilon(10860)\to \pi\pi h_b(mP)$ 
demonstrate a better convergence than those for the $\Upsilon(nS)$ final states.
The analysis of the data on the transitions $\Upsilon(10860)\to \pi\pi\Upsilon(nS)$ with $n=1,2,3$ performed in Ref.~\cite{Baru:2020ywb} relied on the twice subtracted dispersive integrals (\ref{Idisp}), see also Refs.~\cite{Molnar:2019uos,Chen:2016mjn} for related discussions. 
Meanwhile,   since the convergence of similar integrals for the transitions
$\Upsilon(10860)\to \pi\pi h_b(mP)$ with $m=1,2$ is better, it is
sufficient to employ only one subtraction,
\be
\hat{I}_0(s)=
\mbox{const}+\frac{s}{\pi}\int_{4m_\pi^2}^\infty\, ds'\frac{\hat{\Omega}_0^{-1}(s')\hat{T}(s')\hat{\sigma}(s')\hat{M}_0^L(s')}{s'(s'-s-i0)}.
\label{Idisp1}
\ee

Moreover, the form of the function $G_-\propto (m_{z'}^2-m_z^2)$ quoted in Eq.~(\ref{amlsUh}) implies that the corresponding HQSS-breaking subtraction constant can be inferred in the form
\be
\mbox{const}=\frac{m_{z'}^2-m_z^2}{m_{z'}^2+m_z^2}\, c,
\label{c}
\ee
where the mass-dependent factor sets the correct HQSS scaling while the dimensionless constant $c$ is expected to take natural values of the order unity. 

\subsection{Data analysis}
\label{sec:full}

\begin{figure*}[t!]
\centering
\includegraphics[width=0.45\textwidth]{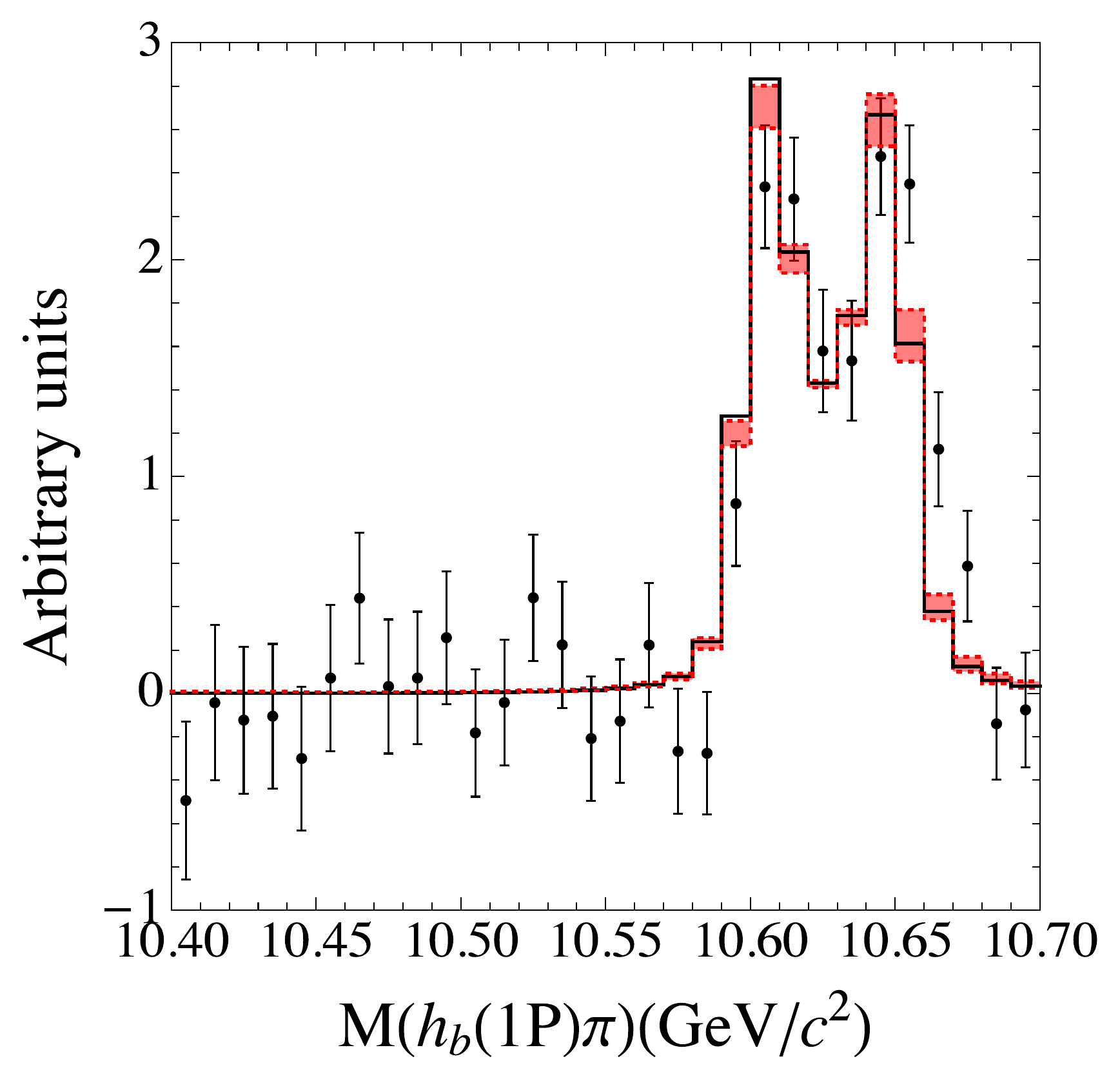} 
\includegraphics[width=0.45\textwidth]{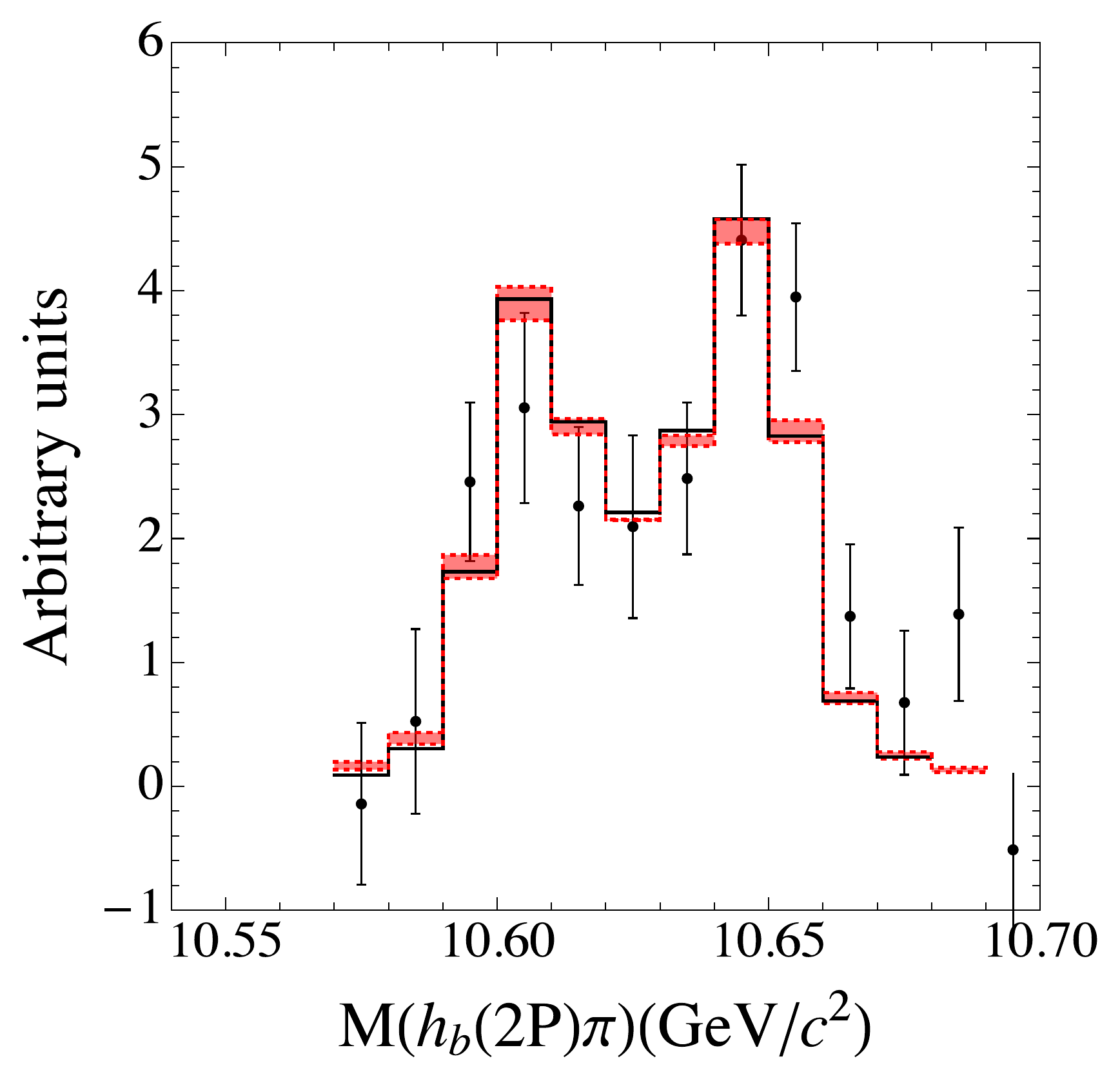}
\caption{Effect of the $\pi\pi/ K\bar K$ FSI on the line shapes $\Upsilon(10860) \to \pi\pi h_b(mP)$ ($m=1,2$).
The solid black line shows the fitted line shapes in the inelastic $h_b(mP)\pi$  channels built in Ref.~\cite{Baru:2020ywb} (fit A)
via the $B^*B^{(*)}$-meson loops but without considering the $\pi\pi/ K\bar K$   FSI. The experimental data are from 
Refs.~\cite{Belle:2011aa,Garmash:2015rfd}. The red band shows the effect of the $\pi\pi/ K\bar K$ FSI on the line shapes when the 
subtraction constant $c$ (see the definition in Eq.~(\ref{c})) is varied  within the interval $[-10,10]$).
 }
\label{fig:lines}
\end{figure*} 

In this section we employ the dispersive approach developed in the previous section to generalise the analysis of the experimental data on the dipion transitions $\Upsilon(10860)\to \pi^+\pi^- h_b(mP)$ with $m=1,2$ performed in Ref.~\cite{Wang:2018jlv}. In particular, we check against the data that the effect of the $\pi\pi$ FSI is small
for a counter term with a strength in line with expectations from heavy quark spin symmetry.

In order to proceed, we stick to the technique of the spectral function previously employed in Ref.~\cite{Baru:2020ywb} to write
\be
M_0^{h_b;L}(s)=\int_{\mu^2_{\rm min}}^{\mu^2_{\rm max}} d\mu^2 \rho(\mu^2) M_{0, \text{stable}}^{h_b;L}(s,\mu),
\label{Eq:M0}
\ee
where the amplitude $M_{0, \text{stable}}^{h_b;L}$ for stable $Z_b$'s
was defined in the previous section while the integration limits are set as in Ref.~\cite{Baru:2020ywb}. The spectral density function $\rho(\mu^2)$ can be evaluated as
\be
\rho(\mu^2)=-\frac{1}{\pi}\mbox{Im\,}U(\mu^2),
\label{rhoFitA}
\ee
with $U(\mu^2)$ referring to 
the production amplitude for $\Upsilon(10860)\to \pi^+\pi^-h_b(mP)$
built in Ref.~\cite{Wang:2018jlv} (fit A) via the $B^*B^{(*)}$-meson
loops but without considering the $\pi\pi$ FSI\footnote{The function
 $U(\mu^2)$ contains also the effect of the finite width of the $Z_b$
 states discussed above.}. Furthermore, since the latter effect is
expected to be small, we do not refit the data but employ all the
parameters found in Ref.~\cite{Wang:2018jlv} from the fit to the data
and only check how the line shapes change upon including 
the $\pi\pi$ FSI and varying 
the HQSS-breaking constant (\ref{c}) within an interval consistent with naturalness. In particular, in Fig.~\ref{fig:lines} we demonstrate the modification of the line shapes in the channels 
$\pi^+\pi^-h_b(mP)$ with $m=1,2$ when the coefficient $c$ in
Eq.~\eqref{c} varies within
a natural interval. 
For definiteness and keeping in mind that the values of $c$ consistent with naturalness should be $|c|\sim 1$, we choose a conservative interval $|c|\leqslant 10$. Comparing the red band, which shows the spread with the parameter $c$ of the line shapes with the $\pi\pi$ FSI included, and the black solid curve, which represents the fit built in Ref.~\cite{Wang:2018jlv}, one can conclude that, in agreement with  the arguments provided in this section, the modification of the line shapes due to the inclusion of the $\pi\pi$ interaction in the final state is marginal.
To be more quantitative, in terms of the $\chi^2/N_{\rm dof}$ this corresponds to a spread from 1.21 to 1.36 as compared with the value 1.29 obtained 
for the fit A from Ref.~\cite{Wang:2018jlv}. 
Moreover, if $c $ is allowed to vary freely, the best fit to the data yields $c(1P) = 24$ and $c(2P) = -24$ with the $\chi^2$/dof =1.16, which is only marginally smaller than the lower bound within the natural interval (the red band in Fig.~\ref{fig:lines}). 
However, such values of the parameter $c$ returned by the best fit are formally somewhat larger than what we regard as natural.

We note also, as a disclaimer, that  
the fit may use the freedom of varying $c$ to partly compensate for other approximations made in the contact fit A of Ref.~\cite{Wang:2018jlv}, for example, the effect of the neglected one-pion exchange.   
Since the resulting effect for the $h_b$ final states is small, we do not dwell on this any further in this work. 
 Although the inclusion of the $\pi\pi/K\bar K$ FSI  in an EFT approach with non-perturbative pions (see, e.g., Ref. [28]) is technically more demanding  because of the additional left-hand cut introduced in this way, this would be an important step in the future more quantitative studies.

\section{Summary}
\label{summary}

In this work we investigate the consequences of heavy-quark spin symmetry and the pattern of its breaking for the measured decays 
$\Upsilon(10860)\to \pi Z_b^{(\prime)} \to \pi\pi\Upsilon(nS)$ ($n=1,2,3$) and $\Upsilon(10860)\to \pi Z_b^{(\prime)} \to \pi\pi h_b(mP)$ ($m=1,2$).
 
These decays appear to have nearly equal probabilities in spite of the fact that a direct transition from $S_{b\bar{b}}=1$ $\Upsilon(10860)$ to $S_{b\bar{b}}=0$ $h_b(mP)$ changes heavy-quark spin and therefore violates HQSS. As argued in Ref.~\cite{Bondar:2011ev}, this can be naturally understood in the molecular scenario for the $Z_b$ states,
because in this case $Z_b(10610)$ and $Z_b(10650)$ appear as orthogonal combinations of the $S_{b\bar{b}}=0$ and $S_{b\bar{b}}=1$ components of equal strength. On the other hand, the interference between the contributions of the $Z_b(10610)$ and $Z_b(10650)$ to the decay amplitude for a $\pi\pi h_b$ final state 
is destructive which results in a vanishing signal in the strict HQSS limit.
In contrast to this, the signal for a $\pi\pi \Upsilon$ final state is
fully in line with HQSS due to a constructive interference between the
two $Z_b$ states. Thus, in order to have signals of a similar strength
in the $\pi\pi \Upsilon$ and $\pi\pi h_b$ final states, HQSS in the
latter channel must be violated ($m_{z'}-m_z\neq 0$), and this violation should be balanced by another quantity with dimension of a mass of a similar (small) size. We demonstrate that this balance is provided by the small widths of the $Z_b$ states $\Gamma_z$. As a consequence, the transition amplitude to the $\pi\pi h_b$ final state strongly depends on the ratio of these two scales, $r=(m_z'-m_z)/\Gamma_z$, as given in Eqs.~\eqref{fm} and \eqref{rdef}. 
Remarkably, while the ratio $r$ vanishes in the strict HQSS limit, its value $r_{\rm phys}\approx 3$ obtained for the physical masses of the $Z_b$'s corresponds to the dynamical regime which is already quite close to the opposite limit of $r\to \infty$ --- in this limit the absolute values of the transition amplitudes to the $\pi\pi \Upsilon$ and $\pi\pi h_b$ final states are shown to have exactly the same peak positions and strengths. 

Also, we pinpoint a crucial difference related to HQSS in the role played by the $\pi\pi/ K\bar K$ FSI in the dipion transitions
$\Upsilon(10860)\to \pi Z_b^{(\prime)} \to \pi\pi\Upsilon(nS)$ and $\Upsilon(10860)\to \pi Z_b^{(\prime)} \to \pi\pi h_b(mP)$.
We demonstrate how the absence of the direct short-range HQSS-conserving operators in the bottomonium transitions $\Upsilon(10860)\ \to \pi\pi h_b(mP)$ can be reconciled with a subtraction in the dispersive integral
if the $\pi\pi/ K\bar K$ FSI is taken into account. In particular, it
is shown that when both $Z_b$ states are included simultaneously,
their leading contributions to the total amplitude cancel against each
other. As a result, the dispersive integrals in the $\pi\pi h_b$
channels demonstrate a better convergence than those in the
$\pi\pi\Upsilon$ channels and the remaining subtraction constants in
the $\pi\pi h_b$ amplitudes are indeed suppressed by HQSS. We have
verified that, in agreement with natural expectations, implementing
this scheme to take into account the $\pi\pi/ K\bar K$ FSI does not
result in any appreciable change in the fitted line shapes in the
$\pi\pi h_b(1P)$ and $\pi\pi h_b(2P)$ channels (contrary to the
crucial role played by the $\pi\pi/ K\bar K$ FSI in the
$\pi\pi\Upsilon(nS)$ channels), so that neglecting the $\pi\pi/ K\bar K$
FSI in the previous analysis of Ref.~\cite{Wang:2018jlv,Baru:2019xnh}
is indeed justified, at least at the present level of accuracy dictated by the quality of the existing data. Meanwhile, our findings establish a way to improve on such an analysis when/if such improvements are finally called for by the experiment. 

{As a final remark, we notice that while the numerical analysis performed in this work relies on a particular fit to the data built in the framework of the molecular approach for the $Z_b$'s, the qualitative arguments presented in this paper are solely based on the interplay of different scales (the mass splitting $m_{z'}-m_z$ and the width $\Gamma_z$ in the first place) 
and therefore can equally apply to any approach to the $Z_b$'s which is consistent with HQSS and  
its leading breaking effects, in particular, to the tetraquark model.}

Our findings concerning the role played by the $\pi\pi/ K\bar K$ FSI should be also important for understanding analogous charmonium transitions to the final states $\pi\pi\psi(nS)$ and $\pi\pi h_c(mP)$ via 
the $Z_c(3900)$ and $Z_c(4020)$ states, though the effects of the HQSS violation in the charmonium sector are expected to be larger.

\begin{acknowledgments}
{The authors thank Sa{\v s}a Prelov{\v s}ek for a fruitful discussion that triggered this project.} This work is supported in part by the NSFC and the Deutsche Forschungsgemeinschaft (DFG) through the funds provided to the Sino-German Collaborative Research Center ``Symmetries and the Emergence of Structure in QCD'' (NSFC Grant No. 11621131001, DFG Grant No. CRC110) and by BMBF (contract No. 05P21PCFP1).  A. N. is supported by the Slovenian
Research Agency (research core Funding No. P1-0035).
\end{acknowledgments}

\end{document}